\documentclass[11pt,a4paper]{article}
\usepackage[nohead, nomarginpar, margin=1in, foot=.25in]{geometry}

\usepackage{graphicx}
\usepackage{amssymb}
\usepackage{caption}
\usepackage{subcaption}

\usepackage[T1]{fontenc}
\usepackage[utf8]{inputenc}
\usepackage{authblk}

\begin{document}
\title{Application of Nilsson model for deformed nucleus in relativistic heavy ion collisions}
\author[1,2]{S. K. Tripathy}
\author[3, \footnote{younus.presi@gmail.com}]{M. Younus}
\author[1]{P. K. Sahu}
\author[2]{Z. Naik}

\affil[1]{Institute of Physics, HBNI, Sachivalaya Marg, Bhubaneswar 751005, India}
\affil[2] {Sambalpur University, Jyoti Vihar, Burla, Sambalpur, Odisha 768019, India}
\affil[3] {Nelson Mandela University, Port Elizabeth, 6031, South Africa}

\maketitle

\begin{abstract}
Electron scattering methods, involving nucleus which have little or no intrinsic deformation suggest nucleon distribution to be of Fermi type. This distribution is further parameterised as Wood Saxon (WS) distribution, where an uniform charge density with smoothed-out surface have been implemented. Incorporating shape modification in WS, earlier attempts were made to explain observables in deformed nuclear collisions, such as charged particle multiplicity.  
In this work, we use an alternate approach known as Nilsson model or Modified Harmonic Oscillator (MHO), to explain charged particle multiplicity in U+U collisions at top RHIC energy. We have implemented the formalism in HIJING model  
and we found that the model describes the experimental data to an extent. 
\end{abstract}

\noindent
\textit{Keywords : Monte Carlo simulations, deformed nuclei, Charged particle production} \\
\textit{PACS: 21.60.Ka, 25.75.Dw} \\


\section{Introduction}
\label{intro}
One of the goals of heavy ion collisions is to produce a system of de-confined quarks and gluons (known as QGP) at high temperatures and densities. 
Existing experimental data and theoretical simulations indicate the formation of collective phenomena at the early stage of QGP. Some of the experimental observable suggests that initial anisotropy in partons' configuration affects the final state particle and hence, corresponding observables.  It is known that both kinematics and dynamics of heavy ion collisions depend upon collision centrality as well as nucleon distributions of colliding nuclei. The muti-scattering of the participant nucleons and their constituent partons give rise to some novel phenomena such as nuclear shadowing, Cronin effects etc \cite{cronin, shadowing}. Similarly, non-participating nucleons or spectators contribute to the chiral magnet effects produced in QGP. So, in order to simulate and study QGP dynamics, initial state geometry plays a very important role and must be precisely evaluated and determined.\\
In case of heavy (little or zero intrinsic deformed) spherical nuclei (approx.  Au, Pb etc) collisions, the standard Wood-Saxon(WS) distribution~\cite{wdsx1} of nucleons inside a nucleus  gives us charged particle multiplicity distributions ($N_{ch}$).
However for collisions of U nuclei, owing to their prolate shape, they can undergo collisions with body-body, tip-tip, or body-tip or with any random configuration and may provide different pre-equilibrium conditions than that described by WS for spherical nuclei collisions. Therefore, Modified Wood-Saxon(MWS)~\cite{mws1} distribution is applied to U+U collision system in order to explain experimental data ~\cite{mws_1,mws_2, mws_3,mws_4,mws_5,mws_6}.
Many in-depth and first-hand information on deformed system could be obtained from these studies using MWS.\\
In this paper, we have taken an alternative approach for calculating $N_{ch}$ distribution. We have started from Nilsson potential or Modified harmonic oscillator(MHO) potential~\cite{nilsson_book,bhaduri_book}, to derive and develop the MHO nucleon distribution for Uranium nucleus and calculate charged particle multiplicity distribution using Glauber formalism within HIJING model ~\cite{hijing}. 

The paper is organised as follow. In Sec \ref{sec:Nilsson_calc} we will present derivation of Nilsson formalism and will explain in detail analytical form of various terms associated with it. In Sec \ref{sec:results_disc} we will show results from this formalism along with Wood Saxon 
We will compare our results along with published data in this section as well. Finally we will summarise in Sec \ref{sec:conclusion}.

\section{Nilsson distribution/Modified harmonic oscillator(MHO)}
\label{sec:Nilsson_calc}
Using semi-classical partition function~\cite{jennings}
\begin{eqnarray} 
Z_{sc}(\beta)=\frac{2}{\hbar^3}\;\int{e^{-\beta H_{sc}}\,d^3r\,d^3p} \,,
\end{eqnarray}
one may derive the the standard Thomas-Fermi relations (Leading Order) for nucleons' single particle distribution $\varrho (r)$, and energy densities inside a nucleus for a given potential $V(r)$ in the Hamiltonian, $H_{sc}=\frac{1}{2}\sum{m_ir_i^2}+V(r) + f(r,p)$ as, 
\begin{eqnarray}
\varrho (r) & =  \frac{1}{3\pi^2}[\frac{2m}{\hbar^2} (\lambda_0 - V(r))]^{3/2} \frac{m^*(r)}{m}
\label{eq:rho_nilsson}
\end{eqnarray}
Here $f(r,p)$ contains interaction terms such as orbital angular momentum, $l$ etc., (~$\displaystyle V_{ll}=-\kappa\mu\hbar\omega_0 (l^2-\langle l^2\rangle)$).The higher order corrections in orders and terms of $\displaystyle\hbar$ have been calculated~\cite{dudek}. Here as a first attempt, we have derived and used the leading order term as shown in~\cite{bengtsson}. Here, V(r ,$\theta$) is the Nilsson potential, 
$\lambda_0$ is the cut on turning point for Nilsson potential, when $\lambda_0 - V(r)$ becomes negative. $m^*(r)$ is the effective mass and can be written as: 
\begin{eqnarray} 
m^*(r) = m/(1 - 2v_{ll} m r^2)
\end{eqnarray}
Where $v_{ll} = \kappa \mu \omega_0 /\hbar$, k and $\mu$ are parameters.

Binomial expansion of Eq. \ref{eq:rho_nilsson} can be written as follows:
 \begin{eqnarray}
\varrho (r) & =  &\frac{1}{3\pi^2}(\frac{2m}{\hbar^2})^{3/2} \lambda_0^{3/2} [1- V(r,\theta)/\lambda_0 ]^{3/2} \frac{m^*(r)}{m} \nonumber \\
 & =  &\frac{1}{3\pi^2}(\frac{2m}{\hbar^2})^{3/2} \lambda_0^{3/2} [1  - (3/2) V(r,\theta)/\lambda_0 \nonumber\\
 & + & (3/8) (V(r,\theta)/\lambda_0)^2 -(3/48) (V(r,\theta)/\lambda_0)^3\nonumber\\ 
 & + & \mathcal{O}((V(r)/\lambda_0)^4)] \frac{m^*(r)}{m}\, [1/fm^{3}]
 \label{eq:density}
\end{eqnarray}
Eq.~\ref{eq:density} is in expanded form assuming $\lambda_0\;>\,V(r,\theta)$. 
Also, the value assumed for $\lambda_0\gtrsim\;V(R_\theta, \theta)$, as $\varrho(r)$ goes smoothly to zero.\\
We have taken the surface radius of the uranium nucleus as \cite{bao}
\begin{eqnarray}
R|_{\theta=0}=R_A\left(1-\frac{2\epsilon_2}{3}\right)\,[fm]\,,\nonumber\\
R|_{\theta=\pi/2}=R_A\left(1+\frac{\epsilon_2}{3}\right)\,[fm]
\end{eqnarray}
where, $\displaystyle R_A(=1.2A^{1/3})$ is the surface radius of undistorted uranium nucleus.\\
The Nilsson form of $V(r, \theta)$ is taken assuming anharmonic oscillator equation for the distorted nucleus (also known as modified harmonic oscillator, MHO) is as follows:
\begin{eqnarray}
\label{pot1}
V(r, \epsilon, \theta) &=& \frac{1}{2}\hbar\omega_0 (\epsilon) \varrho_t^2\;[1+2\epsilon_1 \sqrt\frac{4\pi}{3} Y_{10}(\theta_t)\nonumber\\
 &-&\frac{2}{3}\epsilon_2 \sqrt\frac{4\pi}{5} Y_{20}(\theta_t) + 2\sum_{\lambda =3}^{\lambda_{max}} \epsilon_{\lambda}\sqrt{\frac{4\pi}{2\lambda+1}}Y_{\lambda0}(\theta_t) ] \,,\nonumber\\
\end{eqnarray}
Considering even order terms up to 2nd order as mentioned in Ref. \cite{bengtsson}
 \begin{eqnarray}
 \label{pot2}
V(r, \epsilon, \theta) = \frac{1}{2}\hbar\omega_0 (\epsilon) \varrho_t^2 [1  - \frac{2}{3}\epsilon_2 \sqrt\frac{4\pi}{5} Y_{20}(\theta_t)]\,[MeV]
\label{eq:potential_11}
\end{eqnarray}
where the spherical harmonics,
\begin{eqnarray}
Y_{20}(\theta) &=& \frac{1}{4} \sqrt{\frac{5}{\pi}}(3 \cos^2\theta -1)
\end{eqnarray}
and
\begin{eqnarray} 
\cos\theta_t =  \left[\frac{1-(2/3) \epsilon_2}{1+\epsilon_2[(1/3) - \cos^2\theta]}\right]^{1/2} . \cos\theta
\end{eqnarray}

One can also calculate, \\
$\omega_0 (\epsilon) =   \omega_{00}\,(1 - \frac{1}{3}\epsilon_2^2 - \frac{2}{27} \epsilon_2^3)^{-1/3}$\\
where $\omega_{00}$ value can be calculated from the expression,\\
$\hbar \omega_{00} = 41 \times A^{-1/3}$ MeV (for U, A = 238).

The position of the nucleon from nucleus centre, $\displaystyle\varrho_t$ in the stretched spherical coordinates is given by

 $\varrho_t^2 =  \xi^2 +\eta^2 + \zeta^2\,[fm^2]$\\
where $\displaystyle\xi= x[\frac{m\omega_0(\epsilon)}{\hbar}(1+\frac{1}{3}\epsilon_2)]^{1/2} $, $\displaystyle\eta= y[\frac{m\omega_0(\epsilon)}{\hbar}(1+\frac{1}{3}\epsilon_2)]^{1/2} $ and $\displaystyle\zeta= z[\frac{m\omega_0(\epsilon)}{\hbar}(1-\frac{2}{3}\epsilon_2)]^{1/2} $.\\

Therefore, we have\\
$\displaystyle\Rightarrow \varrho_t^2 =r^2 \frac{m\omega_0(\epsilon)}{\hbar}[\sin^2\theta (1+\frac{1}{3}\epsilon_2)+ \cos^2\theta (1-\frac{2}{3}\epsilon_2)]$ . \\

We have also used quadrupole deformation parameter in our current work~\cite{bengtsson}
\begin{eqnarray} 
\epsilon_2& = &0.944\beta_2 - 0.122 \beta_2^2 + 0.154 \beta_2 \beta_4 - 0.199\beta_4^2
\end{eqnarray}\
Value of $\beta_i$ are mentioned in Table of Ref. \cite{atomic_data_table}. 

Thus, we have Nilsson single particle distribution for nucleons inside Uranium nucleus from Eq.~\ref{eq:density}.

Finally, we have implemented Nilsson distribution in the in-built Glauber formalism of HIJING 
We have taken top RHIC energy (U+U $\sqrt{s_{NN}}$=193 GeV) and charged particles for our calculations. 
In HIJING, density distributions 
generate transverse position of individual nucleons. Between each pair of colliding nucleons, impact parameter is calculated using their transverse positions. Eikonal formalism, which uses straight line trajectories between two nucleons is used to calculate probability of collision.

\section{Results and discussions}
\label{sec:results_disc}

 \begin{figure}
\centering
\begin{subfigure}{.5\textwidth}
  \centering
  \includegraphics[width=\linewidth]{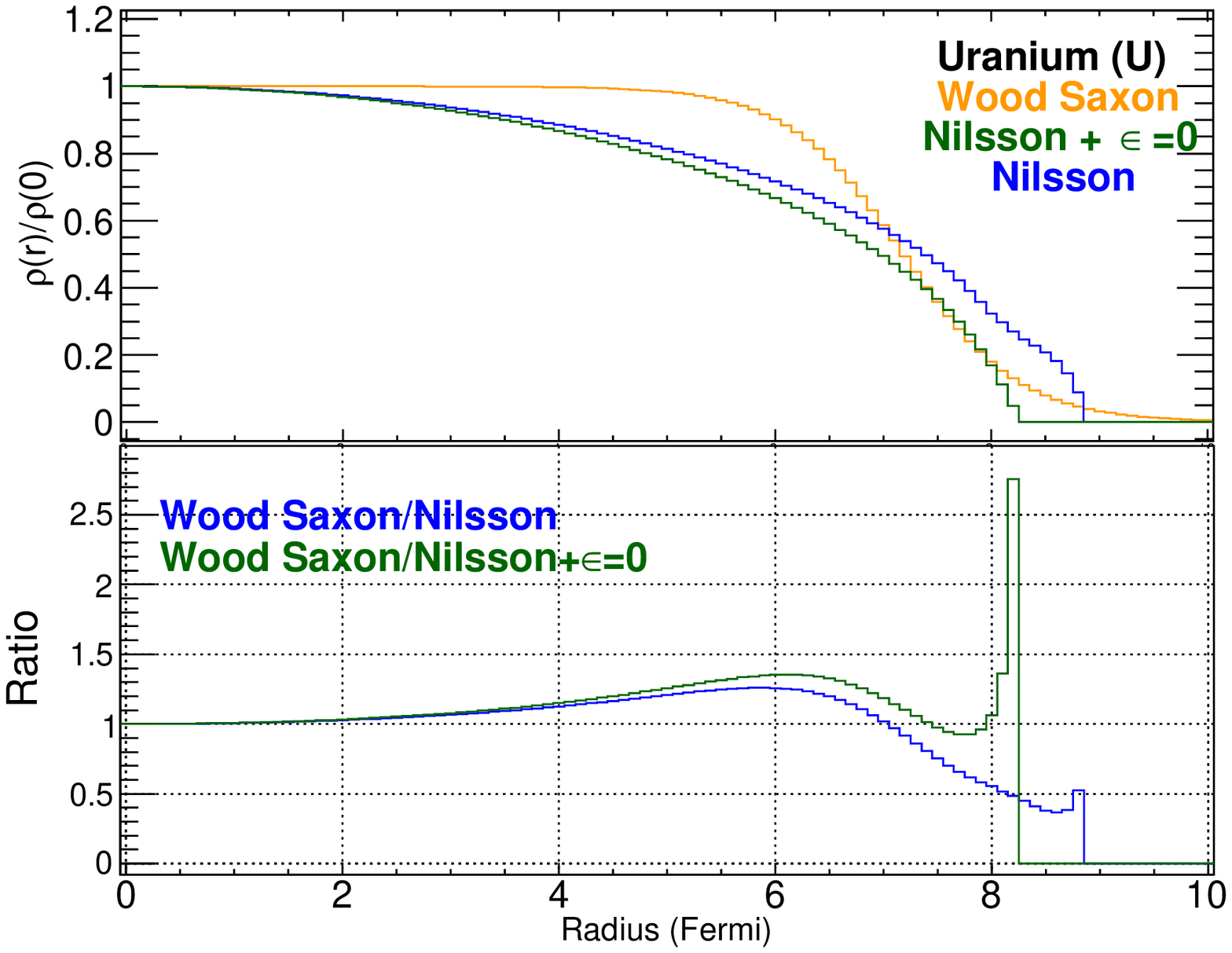} 
    \caption{}
\label{fig:compare_density_normalized}
\end{subfigure}%
\begin{subfigure}{.5\textwidth}
  \centering
 \includegraphics[width=\linewidth]{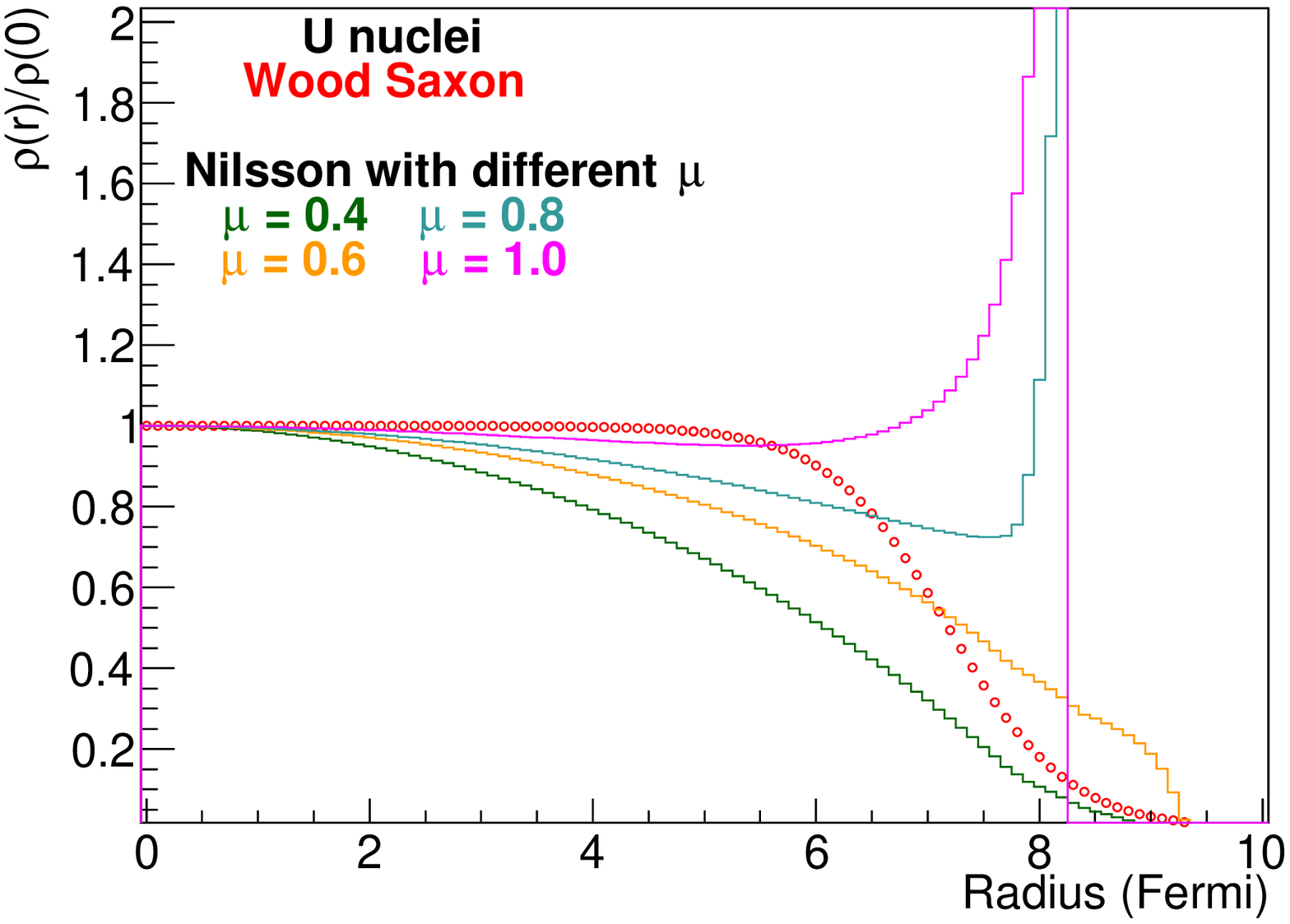} 
   \caption{}
 \label{fig:compare_mu}
 \end{subfigure}
\caption{(Color online)  In the top plot of Fig \ref{fig:compare_density_normalized} density distribution for Wood-Saxon(Orange), Nilsson (Blue) and Nilsson with deformation parameter ($\epsilon$ = 0) in (Green)  in U nuclei are shown. In the bottom plot, ratio between them are shown. In the Fig \ref{fig:compare_mu} It is shown that for one of  $\mu$ values of  Nilsson density distribution, how it converges to Wood Saxon.}
\label{fig:compare_density}
\end{figure}

We have shown normalised nucleon density in Uranium nucleus using Nilsson(MHO)
and standard Wood-Saxon(WS) in Fig. \ref{fig:compare_density_normalized}. Here, we see that distribution from MHO drops rapidly and faster than 
WS. It is in agreement with earlier study by Bengtsson et al.,~\cite{bengtsson} that MHO drops as $\sim e^{-\alpha r^2}$, while 
WS drop as $\sim e^{-\alpha r}$, asymptotically. From $r$= 4 fm onwards, the ratio of MHO shows a deviation from unity.

Earlier results\cite{eAscaatering_1, eAscaatering_2} from electron scattering experiments, suggest a little or zero intrinsic deformations for the nuclei, viz. Ca, V, Co, In, Sb, Au, Bi and C. This investigation also showed that, charge distribution is flat at the central region of the nucleus. Experimental observations however did not include, how the nucleons' distribution would look like for the intrinsic deformed nuclei, such as Hf, Ta, W, U etc.  
On the theoretical front, we know that MWS successfully retains the flat central region even for the deformed nuclei. Our default results using Nilsson density distribution, shows charge density although remains flat in the most central part, starts deviating from central flat region from 2-3 fms onwards.  However, adjusting one of the parameter of Nilsson density (i.e. $\mu$), it goes closer to Wood-Saxon's central flat region even beyond 3 fms as shown in Fig \ref{fig:compare_mu}.

\begin{figure}
\centering
\begin{subfigure}{.5\textwidth}
  \centering
\includegraphics[width=\linewidth]{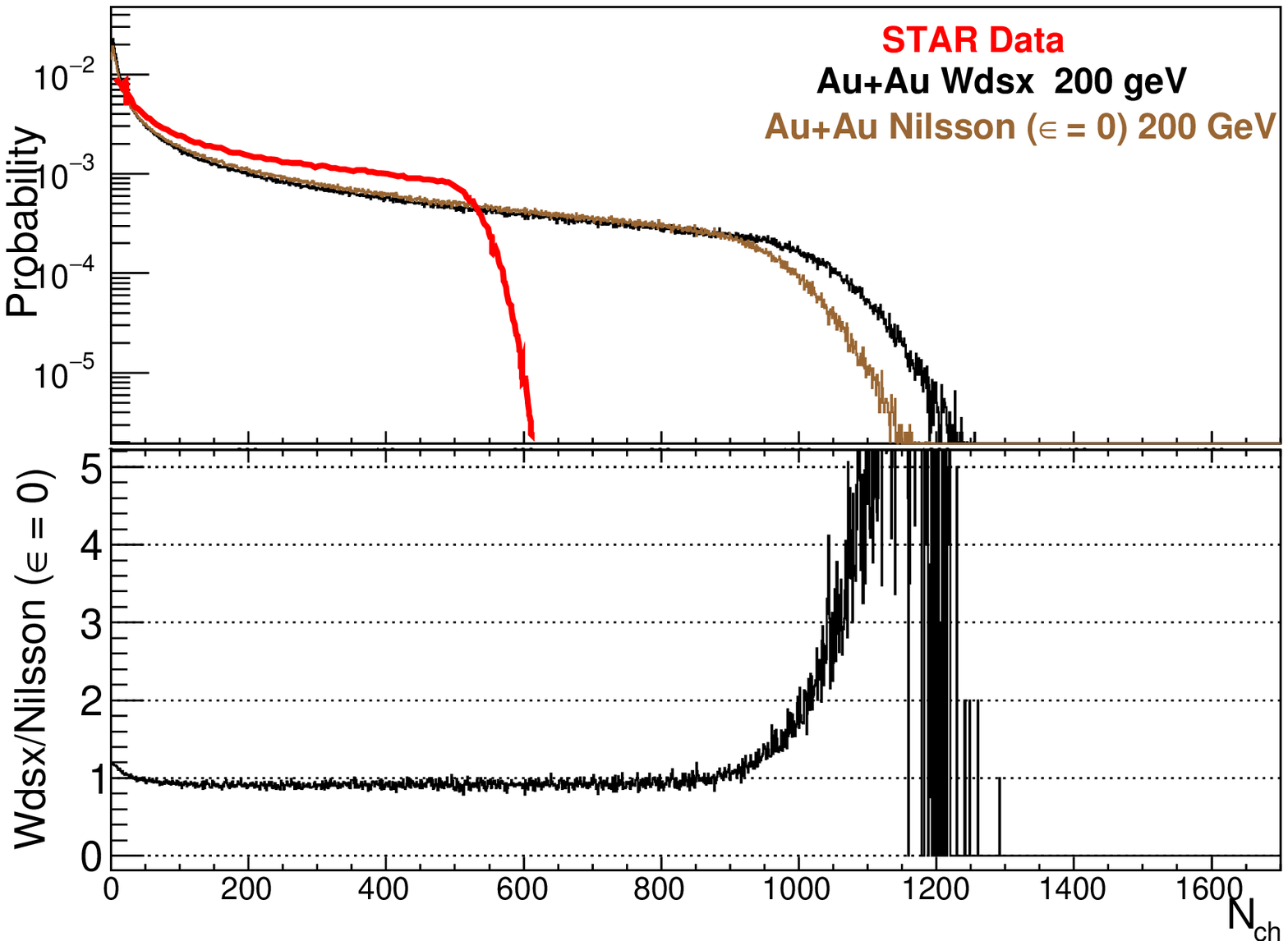}
    \caption{}
\label{fig:compare_nodeformation} 
\end{subfigure}%
\begin{subfigure}{.5\textwidth}
  \centering
   \includegraphics[width=\linewidth]{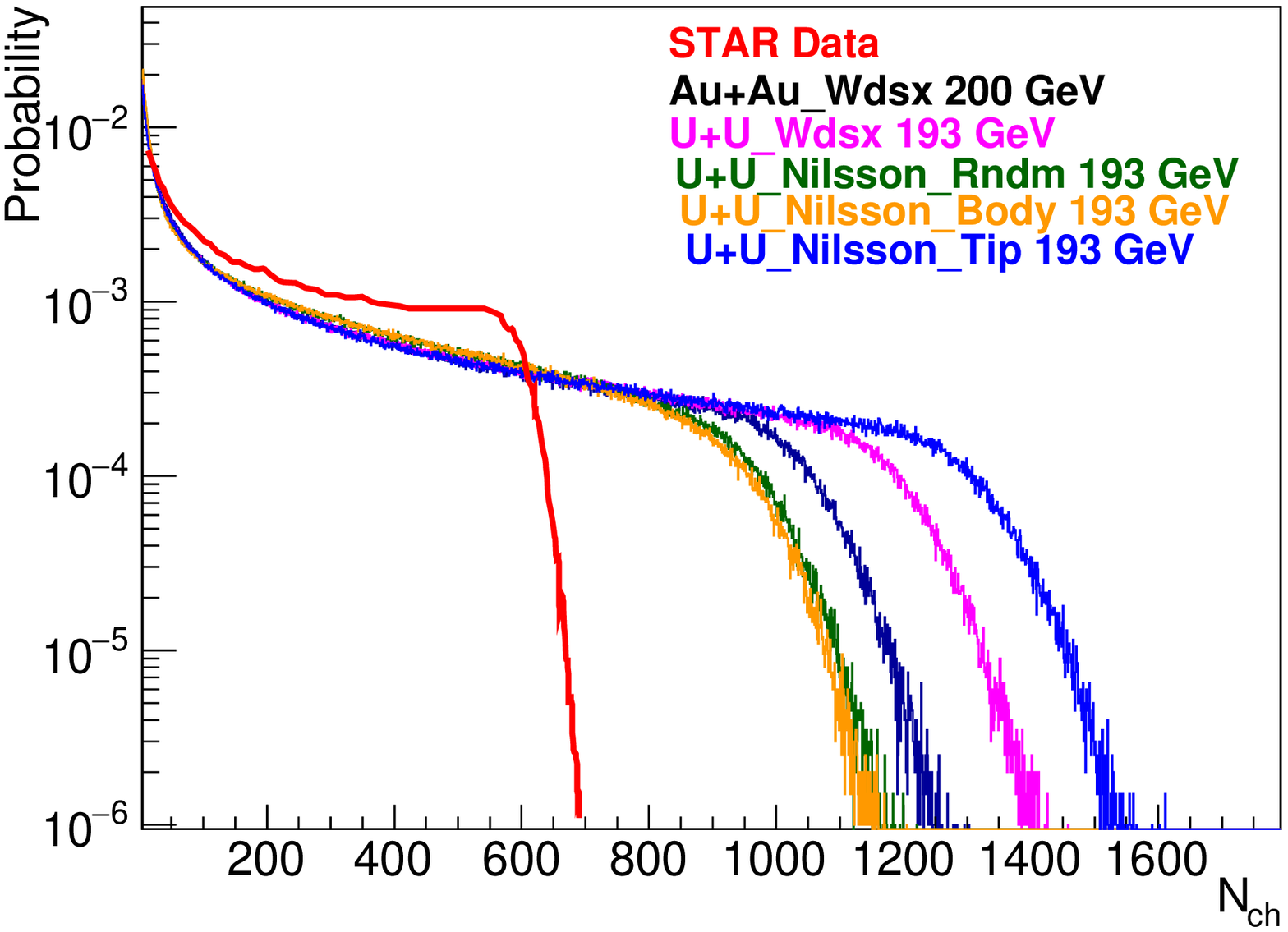}
    \caption{}
  \label{fig:compare_nch_uu}
 \end{subfigure}
\caption{(Color online) In Fig. \ref{fig:compare_nodeformation} HIJING $N_{ch}$ distribution using Wood Saxon and Nilsson (with $\epsilon$ = 0) for Au+Au 200 GeV in the top panel along with experimental data \cite{star_auau_nch}. Ratio of Wood Saxon to Nilsson are shown in bottom panel. In Fig. \ref{fig:compare_nch_uu} HIJING $N_{ch}$ plotted for different configurations of U nuclei along with experimental data \cite{star_uu_nch}. }
\label{fig:compare_nch}
\end{figure}

To investigate further on our formalism on reproducibility of spherical nuclei, we have plotted minimum bias charged particle multiplicity ($N_{ch}$) for Au+Au collision at $\sqrt{s_{NN}}$= 200 GeV using WS and MHO (with $\epsilon$=0) in HIJING model in Fig. \ref{fig:compare_nodeformation}.
WS distribution within HIJING retains the profile shape when compared to experimental results, but gives non-zero charged particle production probability beyond available multiplicity bins from experimental data. On the other hand, MHO with zero deformity reduces the WS estimation in high multiplicity region to some extent, keeping the profile shape almost intact like WS. However, MHO overestimates the experimental data. The plateau region is successfully reproduced by both WS and MHO.

We have plotted our estimation of $N_{ch}$ for U+U collisions at $\sqrt{s_{NN}}$= 193 GeV, with MHO 
and WS formalism from HIJING in Fig \ref{fig:compare_nch_uu}. MHO result is compared to STAR experimental data and keeps the shape intact to an extent, while 
WS show greater multiplicity than MHO. Here we have shown Nilsson results with random angle orientation of target and projectile nuclei(Nilsson -random), with $\theta \ = \ 0$ (Nilsson-tip) and $\theta \ = \ \pi/2$ (Nilsson-body). We observe that, random angle average and body-body configuration yields similar results, while tip-tip configuration gives non-zero probability 
for twice the magnitude of no. of charged particles from random configurations.

\begin{figure}
\centering
\begin{subfigure}{.5\textwidth}
  \centering
\includegraphics[width=\linewidth]{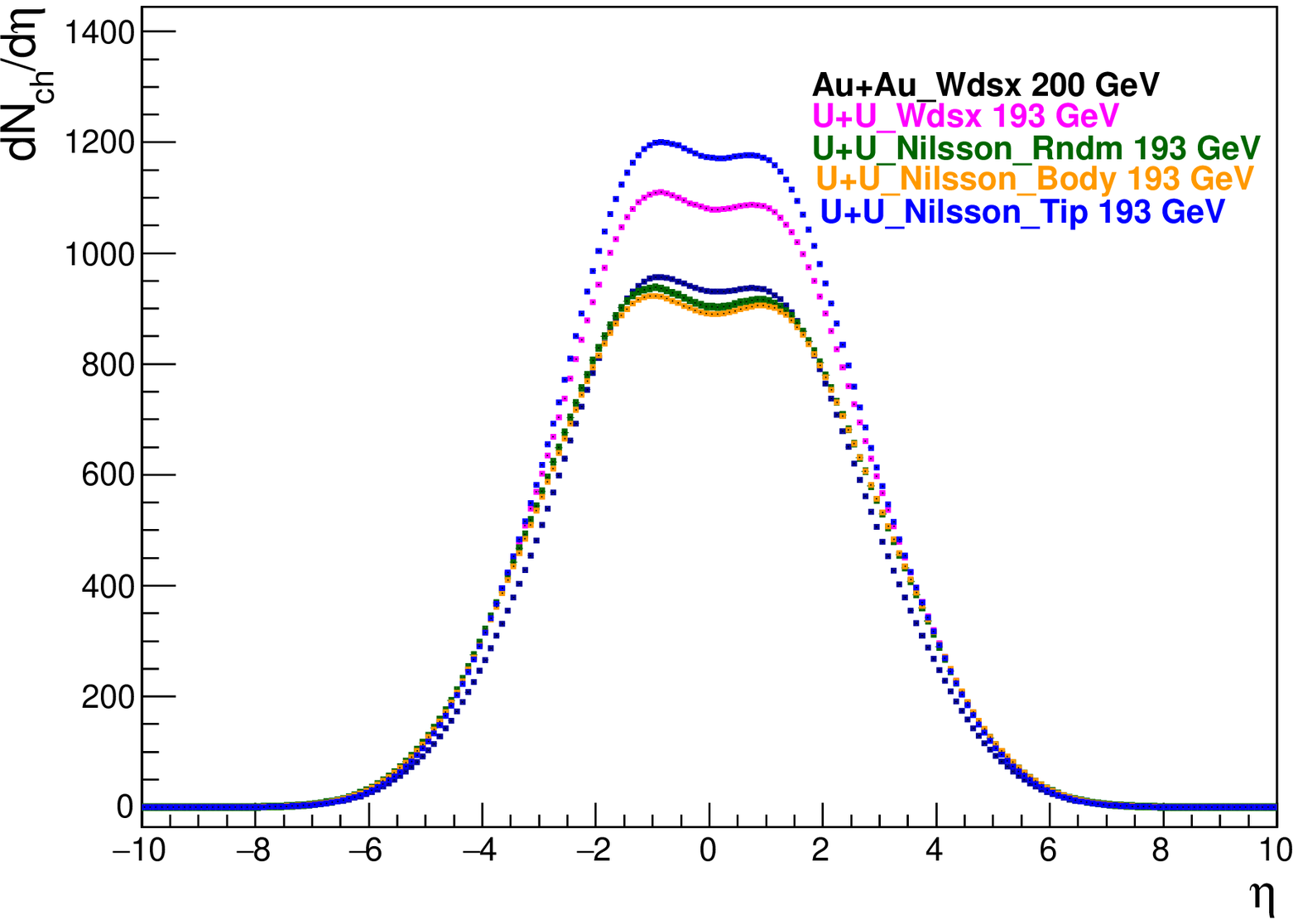}
    \caption{}
 \label{fig:compare_eta_oncent}
 \end{subfigure}%
\begin{subfigure}{.5\textwidth}
  \centering
\includegraphics[width=\linewidth]{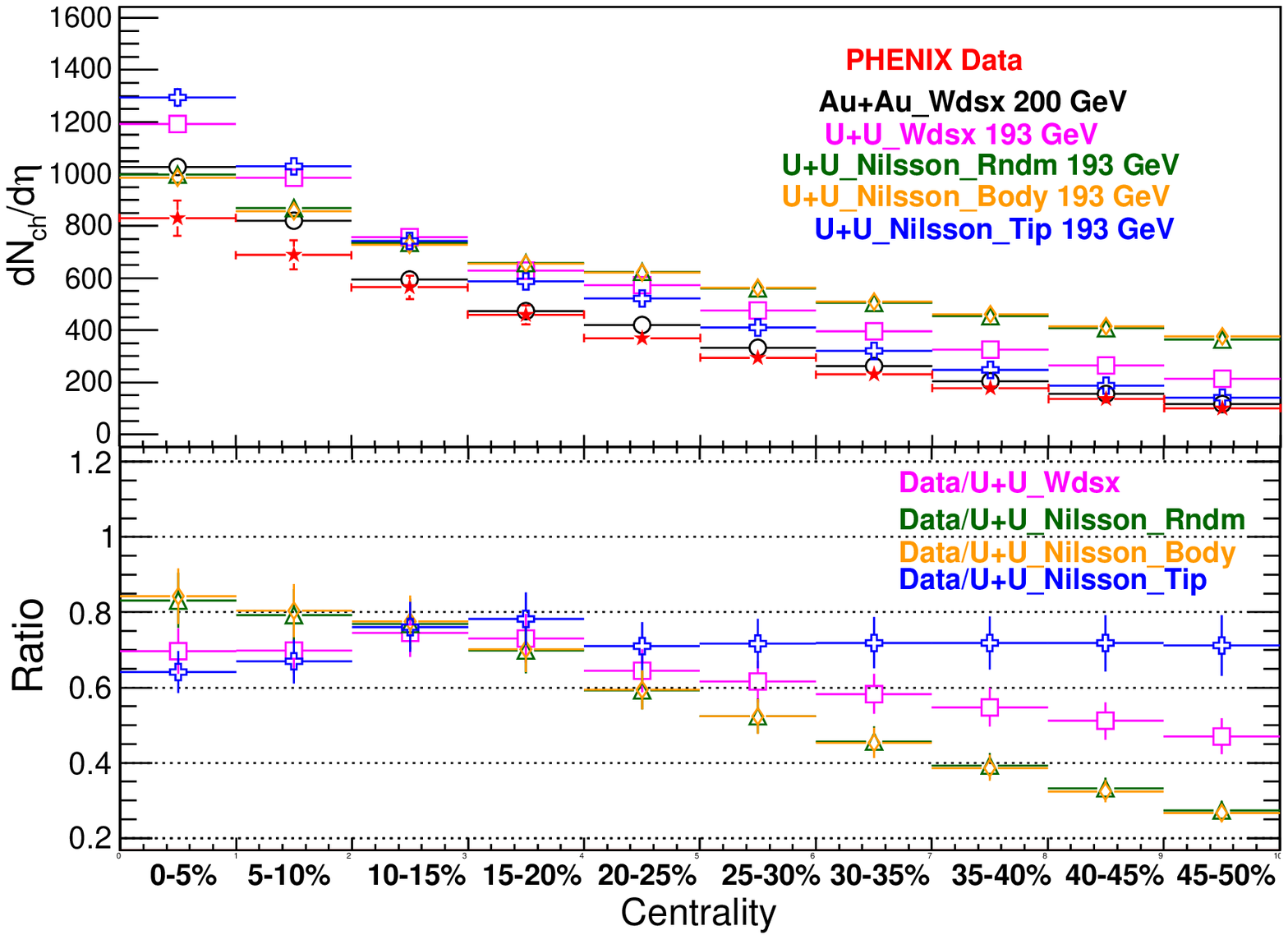} 
    \caption{}
\label{fig:compare_eta_allcent}
 \end{subfigure}
\caption{(Color online) HIJING $dN_{ch}/d\eta$ distribution for central collisions 
($0-5\%$) from WS and MHO (random angle, tip-tip($\theta$ = 0) and body-body 
($\theta \ = \ \pi/2$) ) is shown in Fig \ref{fig:compare_eta_oncent}.  $dN_{ch}/d\eta$  with centrality shown in in Fig \ref{fig:compare_eta_allcent} along experimental data \cite{phenix_uu193}.}
\label{fig:compare_eta}
\end{figure}

We have plotted charged particle pseudo-rapidity distribution ($dN_{ch}/d\eta$) for most central collisions  ($0-5\%$) in U+U collision at $\sqrt{s_{NN}}$= 193 GeV, with MHO 
and WS formalisms from HIJING in Fig \ref{fig:compare_eta_oncent}. MHO gives lower estimates of rapidity distribution compared to WS 
In the similar way of Fig \ref{fig:compare_nch_uu}, we observe in  in Fig \ref{fig:compare_eta_oncent}  that, Nilsson with random angle gives consistent result with body-body configuration but tip-tip configuration yields higher magnitude. 
%
%
%
We have plotted, $dN_{ch}/d\eta$ vs. centrality for U+U collision at $\sqrt{s_{NN}}$ = 193 GeV from HIJING  with WS and MHO formalism in Fig \ref{fig:compare_eta_allcent}. We have compared our results with experimental data\cite{phenix_uu193}. 
Although MHO shows improvement in central collisions regions than WS, our results with MHO for the peripheral collisions divert from WS output. 

\begin{figure}
\centering
\includegraphics[scale=0.5]{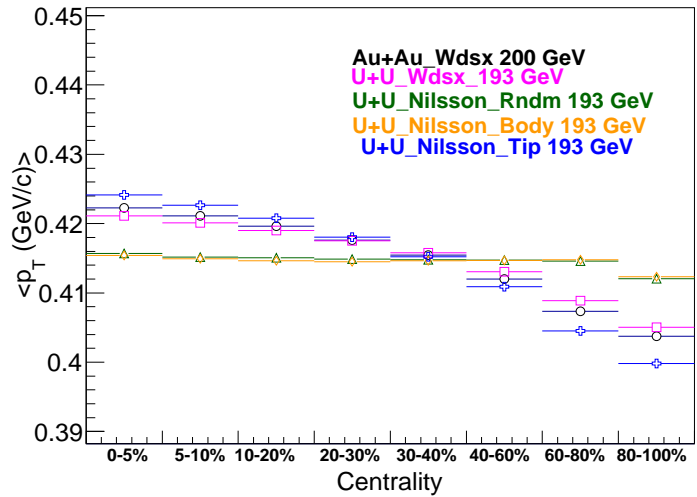}
\caption{(Color online) $<p_{T}>$ distribution of charged hadron from WS and MHO (random angle, tip-tip($\theta$ = 0) and body-body 
($\theta \ = \ \pi/2$) ) }
\label{fig:meanpt} 
\end{figure}

We present average momentum, $<p_T>$, of charged hadrons in Fig. \ref{fig:meanpt}. We observe that body and random configurations of colliding uranium nuclei yield particles 
which show their average momentum independent of centralities of collisions and remain almost flat, 
while other configurations exhibit a downward moving slope with $\approx$ 20\% deviation 
from random configurations. We also observe a reverse in the trend between tip-tip to random configuration, when we go from central to peripheral collision.

\begin{figure}
\centering
\begin{subfigure}{.5\textwidth}
  \centering
\includegraphics[width=\linewidth]{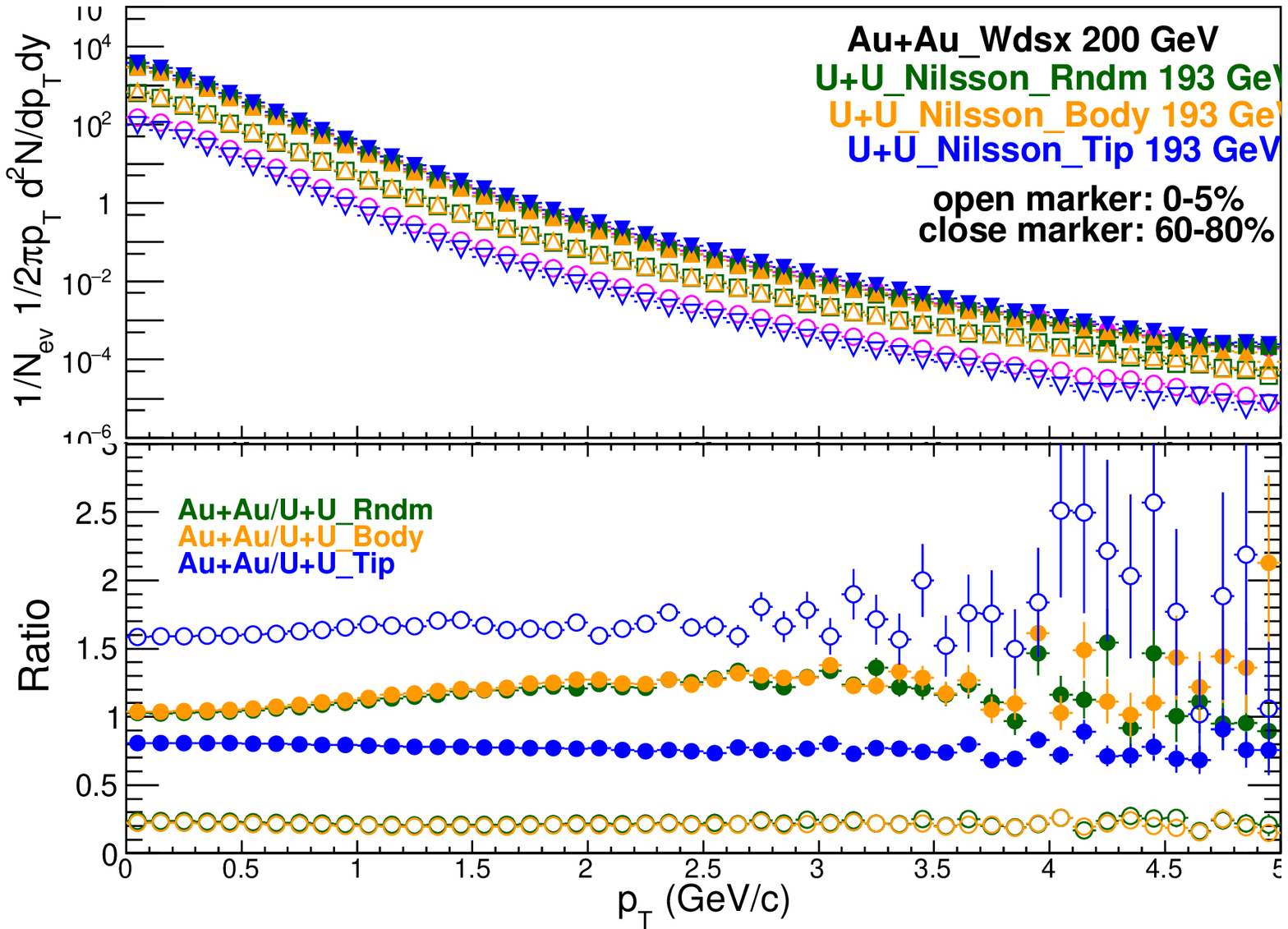}
    \caption{}
 \label{fig:spectra_auau}
 \end{subfigure}%
\begin{subfigure}{.5\textwidth}
  \centering
\includegraphics[width=\linewidth]{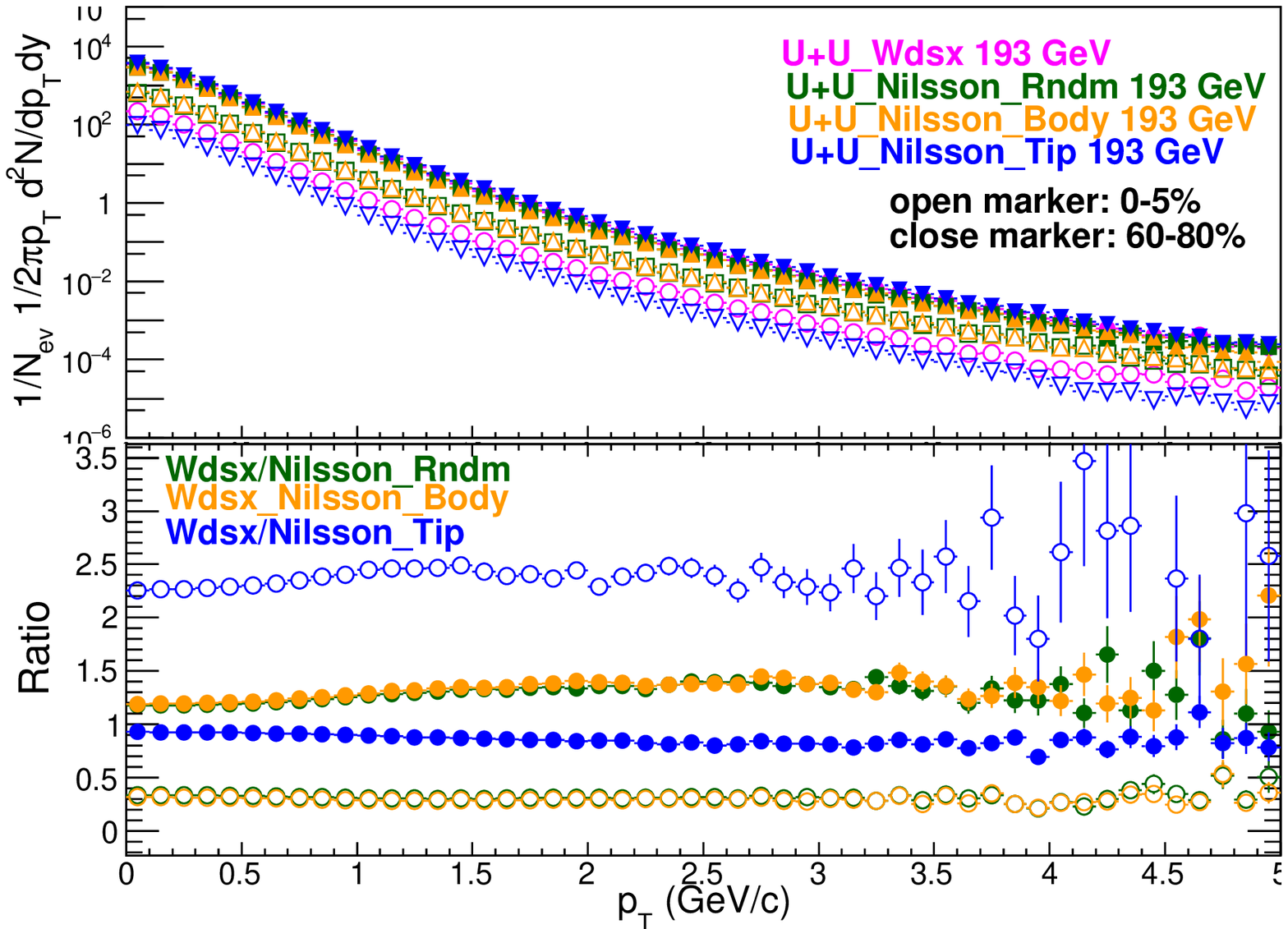} 
    \caption{}
\label{fig:spectra_uu}
 \end{subfigure}
\caption{(Color online) $p_{T}$ spectra of charged hadron in Ws and MHO formalism (with different configurations) in most central and peripheral collision. Results are compared with Au+Au 200 GeV and U+U 193 GeV.}
\label{fig:spectra}
\end{figure}

Transverse momentum spectra of charged hadrons are presented in Fig \ref{fig:spectra}. 
In the top plot of Fig. \ref{fig:spectra_auau}, we have shown our results for most 
central (0-5\%, open markers) and most peripheral (60-80\%,solid markers) collisions in 
U+U collision system alongside Au+Au system. In the bottom plot, 
we have shown the ratios between the various configurations of U+U ($\sqrt{s_{NN}}$ = 193 GeV) 
with Au+Au ($\sqrt{s_{NN}}$ = 200 GeV). 
We observe that, the ratios vary within a broad range of 60\% below unity to 50\% above. 
At both centralities, we find ratios from random configurations match 
with ratios from body-body configurations. 
For the most central collisions, the ratios of particle yields from Au+Au to U+U tip-tip 
configurations differ almost by a large factor of 5. The random or body-body yields seem to 
be always greater than the tip-tip U+U collisions when compared to Au+Au yields.
On the other hand, the tip-tip configuration in peripheral collisions shows lesser value 
than Au+Au with 60\% lower yields, while in the case of body-body or random configurations in 
peripheral 
collisions, we have ratios showing 2\% higher yields than Au+Au. 
We also observe that irrespective of the centralities, the ratios are 
independent of particle transverse momentum, $p_T$. 
In the Fig \ref{fig:spectra_uu}, we calculated the similar ratios, but instead of gold nuclei, 
we have taken uranium nuclei assuming zero deformity. 
The collisions energy is taken to be $\sqrt{s_{NN}}$ = 193 GeV, for all the systems in this case. 
Here too we find similar trends as in previous plot but the variations in the ratios 
have doubled for central collisions as compared to ratios with Au+Au.

\begin{figure}
\centering
\begin{subfigure}{.5\textwidth}
  \centering
\includegraphics[width=\linewidth]{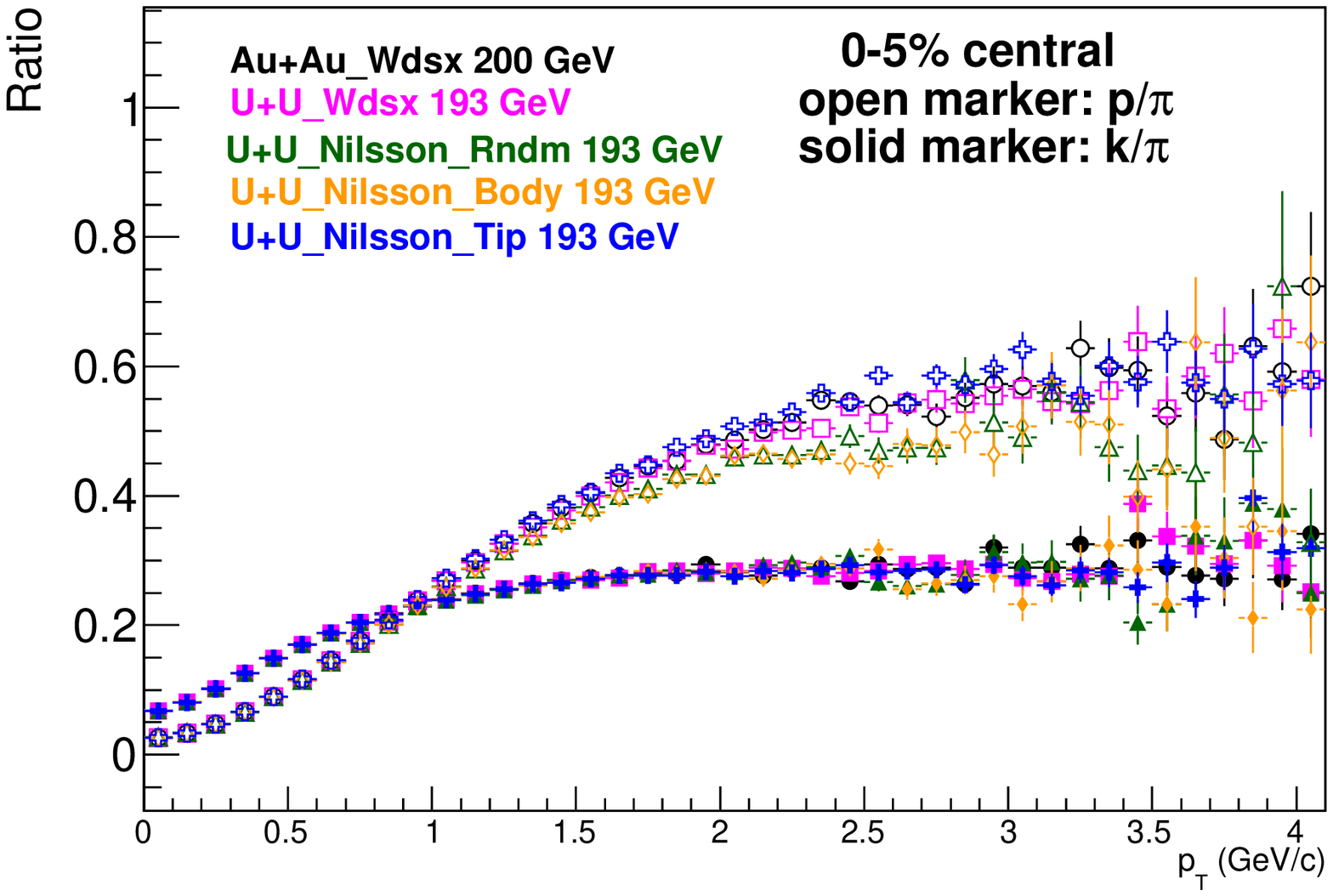}
    \caption{}
 \label{fig:ratio_1}
 \end{subfigure}%
\begin{subfigure}{.5\textwidth}
  \centering
\includegraphics[width=\linewidth]{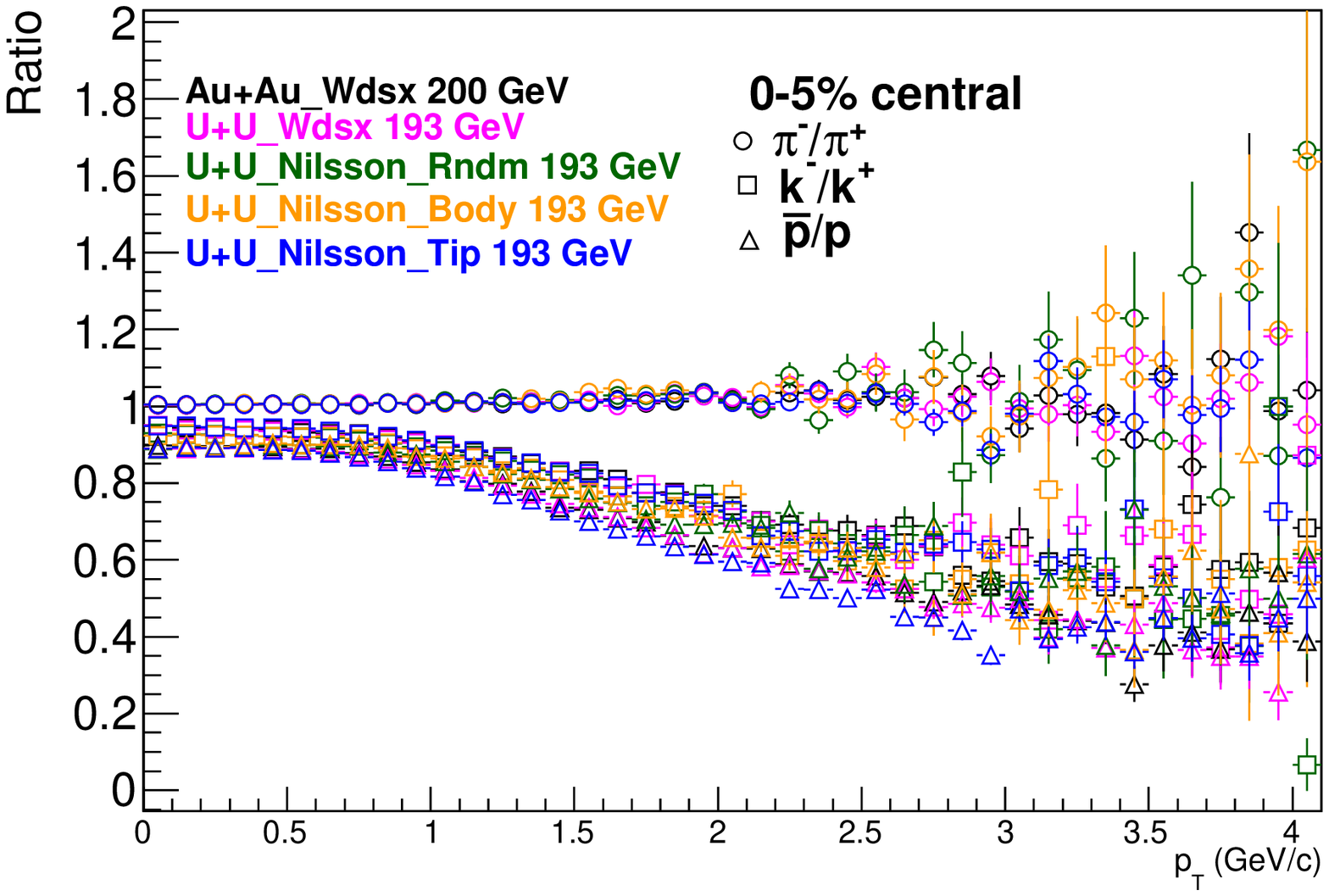} 
    \caption{}
\label{fig:ratio_2}
 \end{subfigure}
\caption{(Color online) particle ratio in most central collision in Ws and MHO formalism (with different configurations).}
\label{fig:particle_ratio}
\end{figure}

We have presented charged hadrons ratios in Fig \ref{fig:particle_ratio} for the 
most central collisions. $p/\pi$ and $k/\pi$ ratios are presented in 
Fig \ref{fig:ratio_1}, while anti-particle to particle ratios are 
presented in Fig. \ref{fig:ratio_2}. We don't find any 
configuration dependencies in any of these ratio plots. There is no
variation of results observed, when two different systems (Au+Au and U+U) are taken. We guess, although 
uranium is heavier than gold nucleus as well as more deformed, the similarities in 
the particle ratios demonstrate their dependencies on the collision energies rather than 
system sizes. 
We also find that around  $p_T$= 1 GeV, $p/\pi$ goes higher from 
$k/\pi$ ratio. $\pi^-/\pi^+$ remains 
almost independent of $p_T$ and value of ratio is around unity. 
However, $k^-/k^+$ and $\bar{p}/p$ ratios decrease from unity with increasing $p_T$.

\section{Conclusion}
\label{sec:conclusion}

We have implemented and tested Nilsson or Modified Harmonic Oscillator(MHO) distribution, to explain some of the observed experimental results for U+U collisions at available RHIC energy within HIJING code. 
Along with this study, we have shown results from Wood Saxon 
and compared with available experimental data. At the first attempt, MHO gives good estimation to experimental charged particle multiplicity distribution as well as to Wood-Saxon formalisms in central collisions.
Without any deformation, MHO also reproduces the plateau region of $N_{ch}$ distribution approximately for gold (Au) nuclei collisions, assumed to be spherically symmetric in our present calculations.

In this study we have taken random angle orientation of U nuclei along beam axis as well as two specific orientations, i.e. body-body and tip-tip. Our study shows that within HIJING formalism, tip-tip orientations of colliding nuclei can generate significantly larger number of particles than body-body or orientation average configurations. We observe that body-body configuration gives similar magnitude with that of orientation average.



\end{document}